\documentclass[preprints,article,accept,moreauthors,pdftex]{myarxiv}

\hreflink{https://doi.org/} 
\usepackage{physics}
\usepackage{siunitx}

\usepackage{graphicx} 
\usepackage{algorithm}
\usepackage{algpseudocode}

\newcommand*{\D}{\mathrm{d}}
\renewcommand*{\eqref}[2][Equation]{\mbox{#1\ (\ref{eq:#2})}}
\newcommand*{\figref}[2][Figure]{\mbox{#1\ \ref{fig:#2}}}
\newcommand*{\secref}[2][Section]{\mbox{#1\ \ref{sec:#2}}}

\Title{Sampling quantum states with inequality constraints}
\Author{%
  Weijun Li,$^{1}$ %
  Rui Han,$^{2}$ %
  Jiangwei Shang,$^{3}$ %
  Hui Khoon Ng,$^{4,5}$ %
  Berthold-Georg Englert$^{\,3,4,2}$}

\address{%
$^{1}$ Department of Physics, University of Oxford, Oxford OX1 3RH, %
            United Kingdom\\
$^{2}$ BiQut, Singapore 288564, Singapore\\
$^{3}$ School of Physics, Beijing Institute of Technology, %
            Beijing 100081, China\\
$^{4}$ Department of Physics, National University of Singapore, %
            Singapore 117542, Singapore\\
$^{5}$ Centre  for  Quantum  Technologies, Singapore 117543, Singapore\\
Email addresses: liweijun2718@outlook.com, %
                 han.rui@quantumlah.org, %
                 jiangwei.shang@bit.edu.cn, \\ \phantom{Email addresses:}
                 cqtnhk@nus.edu.sg, %
                 berge@bit.edu.cn}

\abstract{%
Random samples of quantum states with specific properties are useful for
various applications, such as Monte Carlo integration over the state space.
In the high-dimensional situations that one encounters already for a few
qubits, the quantum state space has a very complicated boundary, and it is
challenging to incorporate the specific properties into the sampling algorithm.
In this paper, we present the Sequentially Constrained Monte Carlo (SCMC)
algorithm as a powerful and versatile method for sampling quantum states in
accordance with any desired properties that can be stated as inequalities.
We apply the SCMC algorithm to the generation of samples of bound entangled
states; for example, we obtain nearly ten thousand bound entangled two-qutrit
states in a few minutes---a colossal speed-up over independence sampling, which
yields less than ten such states per day.
In the second application, we draw samples of high-dimensional quantum states
from a narrowly peaked target distribution and observe that SCMC sampling
remains efficient as the dimension grows.
In yet another application, the SCMC algorithm produces uniformly distributed
quantum states in regions bounded by values of the problem-specific target
distribution; such samples are needed when estimating parameters from the
probabilistic data acquired in quantum experiments.
}

\begin{document}

\section{Introduction}
Random samples of quantum states with specific probability distributions
and/or governed by specific physical properties are very useful for many
applications in quantum information science and other research areas.
For instance, they play an important role in quantum state and parameter
estimation~\cite{PhysRevA.55.R1561, PhysRevA.92.012108,
  PhysRevLett.117.010503}, allowing one to test and verify properties for
certain classes of quantum states~\cite{PhysRevLett.100.110502,
  PhysRevLett.109.040502, 6967820}, as well as to test quantum channels and
processes, such as the random quantum circuits leading to Quantum
Supremacy~\cite{Lund2017, Bouland2019, shortGoogleQSupremacy}.
Similar to sampling classical systems, independent rejection sampling methods
are often inapplicable to high-dimensional quantum systems as they severely
suffer from the exponential growth of resource requirements, known as the
``curse of dimensionality"~\cite{Donoho2000}.
The positivity boundaries of the quantum state spaces, moreover, are typically
extremely complex to characterize for otherwise convenient parameterizations,
causing quantum states to be more challenging to sample.  
 
Most existing methods for the direct sampling of the random density matrices,
which represent the quantum states numerically, rely on either clever
parameterization of the state space or various kinds of induced
measures~\cite{Wootters1990, Bengtsson2006, AlOsipov2010, Zyczkowski2011,
  Collins2016}.
The major drawback is that the exact probability distribution for random
states constructed this way are usually not known; one has to estimate
approximate distributions, and doing this reliably is a numerical challenge
in itself.
Thus, rejection sampling methods that proceed from known proposal
distributions are still often favored for sampling from a desired target
distribution. 
However, even if one can directly draw from a reliable reference distribution
that well mimics the target distribution---for example, using a version of the
complex Wishart distribution
\cite{Wishart1928,Goodman1963a,Goodman1963b}---the exponential
reduction of the acceptance rate as the dimension increases makes it rather
impractical to go beyond sampling small quantum systems (see, for example,
Reference~\cite{Han2021Wishart}).%
\footnote{Other rejection-sampling-based methods, such as adaptive rejection
  sampling, suffer from this problem, too.}

Monte-Carlo-based algorithms~\cite{Hastings1970} can be much more resilient
against this curse of dimensionality.
The major difficulty encountered in using Monte Carlo (MC) methods comes from
the incorporation of constraints into the parameter space.
For instance, enforcing the positivity constraint needed for a physical state,
which is intrinsic to quantum systems, can be extremely challenging. 
Algorithms that rely on clever random walks in the probability simplex, such
as the Markov Chain Monte Carlo (MCMC), suffer from low acceptance rates
because of this positivity constraint, and the Hamiltonian Monte Carlo (HMC)
algorithm~\cite{neal2012mcmc}, whose random walk stays in the state space,
depends on the evaluation of a Jacobian determinant and its derivatives, found
to be numerically unreliable in a high-dimensional quantum state
space~\cite{Shang2015, Seah2015}.

In this work, aiming at overcoming the curse of dimensionality as well as at
efficiently enforcing quantum constraints,  we propose the use of the
sequentially constrained Monte Carlo (SCMC) samplers for the sampling of
quantum states in accordance with a given distribution and/or certain physical
properties.  
The SCMC samplers were first proposed by Golchi and Campbell in
2016~\cite{GOLCHI201698} to effectively impose constraints when sampling
classical systems as an extension to the sequential Monte Carlo (SMC) samplers
proposed by \mbox{Del Moral \emph{et al.}} a decade earlier~\cite{oSMC}. 
SMC methods~\cite{Doucet2001, Chopin2002}, which have been used extensively in
the context of sequential Bayesian inference, are not new to the field of
quantum information.  
For example, they were well used in adaptive/online Bayesian Hamiltonian
estimation~\cite{smcParaEst1, PhysRevA.85.052120, PhysRevLett.117.010503}.  
In the context of sequential Bayesian inference, the sequence of distributions
is constructed as more and more measurement data are taken into consideration.  
Different from SMC methods in Bayesian inference, the SMC samplers enable
efficient sampling using MCMC algorithms through the introduction of a
sequence of artificial intermediate distributions that bridge between an
easy-to-sample initial distribution and the difficult-to-sample target
distribution.  
One application of the SMC samplers in quantum information science was
recently presented in~\cite{PhysRevA.96.052306}.  
Here, we employ the SMC samplers to sample quantum states making use of the
SCMC samplers to better incorporate the constraints that naturally arise in
quantum problems. 

We demonstrate the potential of the SCMC samplers in sampling
quantum states through three concrete examples, which are otherwise difficult,
or even impractical, to achieve using other existing methods.
In the first example, we show how the SCMC samplers can be applied to sample
quantum states with bound entanglement through the imposition of soft
constraints.
Large samples of uncorrelated bound entangled bipartite systems with
dimensions $3\times3$, $3\times4$, $4\times4$, and $3\times5$ are
reported, and a curious property of the $2\times4$ system is
observed.
In this context, we note the SCMC algorithm does not rely on a particular
parameterization of the bound entangled state; see \secref{BE}.

The second example presents a generic approach to sampling quantum states from
desired target distributions.
The efficiency and reliability of our algorithm is demonstrated via sampling
three-qubit and four-qubit states.
Moreover, the results indicate that the computational time scales polynomially
rather than exponentially with respect to the system's dimension.

In the last example, we sample uniformly distributed quantum states in regions
bounded by contours of the target distribution---uniform with respect to the
volume element induces by the Hilbert--Schmidt distance, that is.  
We improve on the implementation of the method recently introduced by Oh,
Teo, and Jeong (OTJ) \cite{PhysRevLett.123.040602, PhysRevA.100.012345} and so
confirm that the direct SCMC sampling from the target distribution is
reliable.
In addition, we observe that the direct SCMC sampling is more efficient than
the indirect OTJ method.
Depending on one's point of view, the last example can be regarded as
benchmarking SCMC against OTJ or OTJ against SCMC.

Some technical details are reported in the Appendix.
Selected samples and codes can be fetched from a dedicated
repository~\cite{SCMCrepository}.

\section{SCMC sampling}
The SMC samplers presented in Reference~\cite{oSMC} enable one to sample
$\mathbf{x}$ efficiently from its difficult-to-sample target distribution
$f(\mathbf{x})$ through a sequence of intermediate distributions.
We choose an initial distribution $g(\mathbf{x})$, which can be the prior or
any appropriate reference distribution that is easy to sample from directly,
and find a discrete sequence of $N_\tau$ density distributions
$\big(h_i(\mathbf{x})\big)_{i=0}^{N_\tau}$ that smoothly bridges between
$g(\mathbf{x})$ and $f(\mathbf{x})$.
One particular choice of $h_i(\mathbf{x})$ is to follow the geometric
path~\cite{Gelman1998, Neal2001},  
\begin{equation}\label{eq:hgen}
  h_i(\mathbf{x})=f(\mathbf{x})^{\tau_i}g(\mathbf{x})^{(1-\tau_i)}\,,
\end{equation}
where $\tau_i$ runs from $0$ to $1$ in arithmetic progression, i.e.,
$\tau_i=i/N_\tau$ as $i$ goes from 0 to $N_\tau$.
Thus, ${h_0(\mathbf{x})=g(\mathbf {x})}$ and
${h_{N_\tau}(\mathbf{x})=f(\mathbf{x})}$.

The utility of SMC samplers was extended by Golchi and
Campbell~\cite{GOLCHI201698} to SCMC samplers; in their approach the hard
constraints on the parameter space are gradually incorporated into the
intermediate distributions as soft probabilistic constraints.
Here, soft constraints refer to relaxations that approach the hard constraint
in the limit. 
For instance, when the desired hard constraint is ${\kappa(\mathbf{x})>0}$
with its indicator function being the step function
${I_\kappa(\mathbf{x})=\eta\bigl(\kappa(\mathbf{x})\bigr)}$, the intermediate
indicator functions for the soft constraints can be smooth approximations of
the Heaviside step function, such as 
\begin{equation}\label{eq:constrgen}
  I_{\kappa,i}(\mathbf{x})
  =\frac{1+\tanh\bigl(a\tau_i\kappa(\mathbf{x})\bigr)}{2}\,,
\end{equation}
where the \emph{tolerance} $a$ is an adjustable scale controlling the
\emph{hardness} $a\tau_i$ of the constraint.
In this case, the indicator function $I_{\kappa,i}(\mathbf{x})$ is
incorporated into the SCMC sampler by taking 
\begin{equation}\label{eq:hgen2}
  h_i(\mathbf{x}) \to h_i(\mathbf{x})I_{\kappa,i}(\mathbf{x})\,,
\end{equation}
and, in the limit of $a\tau_i\to\infty$, it converges to the hard constraint,
i.e.,  
\begin{equation}
  I_{\kappa,N_\tau}(\mathbf{x})\to
  \left\{\begin{array}{l@{\enskip\textrm{for}\enskip}l}
           1&\kappa(\mathbf{x})>0\,,\\
           0&\kappa(\mathbf{x}) < 0\,.\end{array}\right.
\end{equation}
In practice, for a large but yet finite value of $a$, points with
${\kappa(\mathbf {x})\leq0}$ (${\kappa(\mathbf{x})>0}$) are  rejected (accepted)
with a higher and higher probability as ${\tau_i\to1}$, thus sample points
gradually move towards the region satisfying the constraint.
At the final step of the SCMC algorithm, we impose the hard constraint instead
of the probabilistic constraint and reject the sample points that violate the
constraint.
Consult the Appendix for implementation details.

The ability to enforce constraints via SCMC is very useful for sampling
quantum states, because the quantum Hilbert space intrinsically requires the
physicality constraint which usually means complicated constraints on the
parameters (positivity of a large matrix) that are CPU-expensive to check.
Apart from the physicality constraint, many other interesting quantum
constraints can also be extremely difficult to enforce as suitable
parameterizations of the states are unavailable.
For efficient sampling of quantum states, we need to identify the quantum
constraints and find their corresponding indicator functions.
Next, we show how well the SCMC sampler works for sampling quantum states in
three distinct contexts.

\section{Examples}
\subsection{Bound entanglement}\label{sec:BE}
\begin{figure*}[!t]
  \centerline{\includegraphics[viewport=50 425 545 755,clip]%
  {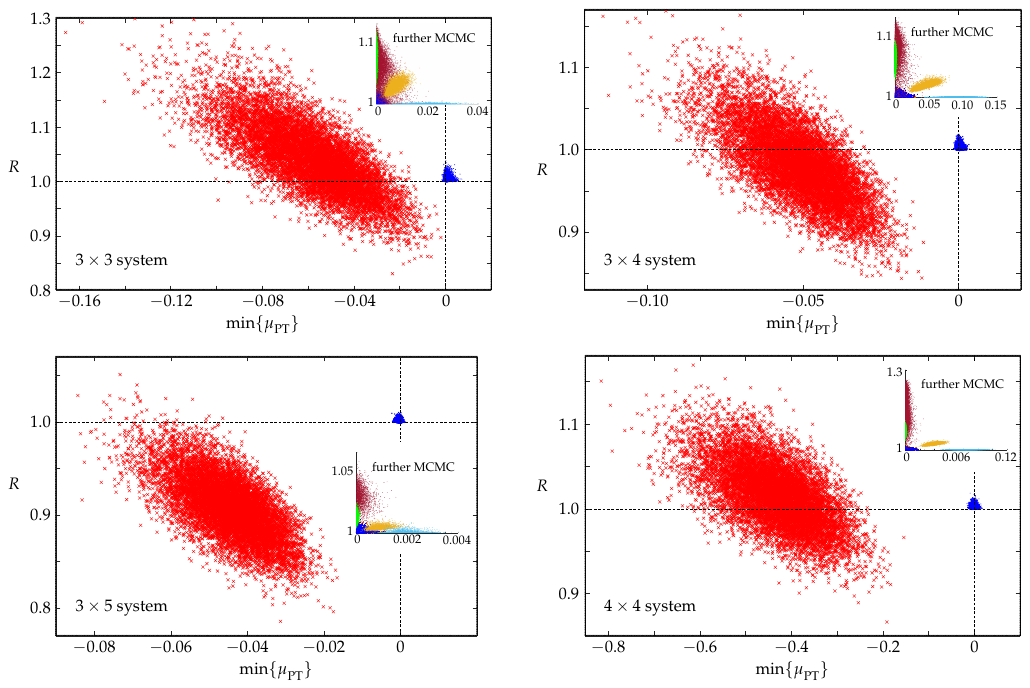}}%
\caption{\label{fig:BE1}%
  The generation of bound entangled states using SCMC. The initial
  reference sample has $10^4$ states (red crosses) uniformly distributed with
  respect to the Hilbert--Schmidt distance shown here on an $R$ vs.\
  $\min\{\mu^{\ }_{\mathrm{PT}}\}$ plot.
  The states after SCMC are indicated by the blue dots. Out of these 10,000
  post-SCMC states, 8,530, 7,011, 2,211, and 4,013 states are bound
  entangled for the respective systems of dimensions $3\times3$, $3\times4$,
  $3\times5$, and $4\times4$. 
  The insets show samples filtered through further MCMC iterations
  with different propagation kernels (by choosing different directions of the
  random walks) represented by dots of different colors.
  }  
\end{figure*}

A bound entangled state is a state with non-distillable entanglement.
It has drawn a lot of attention in the quantum information community since the
prediction of its existence in 1998~\cite{PhysRevLett.80.5239}; see
\cite{Hiesmayr+2:25} for a recent review.
Random samples of bound entangled states are useful not only for the field of
quantum information but also for mathematical interests.
On the one hand, states with bound entanglement provide a testbed for
studying the relation between entanglement, steering, and Bell-type
nonlocality~\cite{PhysRevLett.113.050404, PhysRevLett.124.050401}; on the
other hand, bound entangled states enable the study of positive maps from a
different perspective~\cite{HORODECKI19961}. 
Moreover, despite being highly mixed, such states are potentially useful for
quantum cryptography and quantum
metrology~\cite{PhysRevLett.94.200501,Czekaj2015}.  

While it is currently unknown if all bound entangled states have a
positive partial transpose (PPT), we do know that all entangled PPT states are
bound entangled~\cite{PhysRevLett.80.5239}, and a PPT state is surely
entangled if the computable cross norm or realignment (CCNR) criterion is met
\cite{CCNR}. 
Taken together, then, the two conditions
\begin{equation}
  \min\{\mu^{\ }_{\mathrm{PT}}\}\ge0\quad\mathrm{and}\quad
  \sum_{j}\sigma_j(\tilde{\rho})\equiv R>1
\label{eq:BEcriteria}
\end{equation}
are sufficient to ensure bound entanglement.
Here, $\{\mu^{\ }_{\mathrm{PT}}\}$ is the set of eigenvalues of the partial
transpose of the density matrix $\rho$ and $R$ is the sum of the singular
values of the realigned matrix $\tilde\rho$ \cite[Eq.~(5)]{CCNR}. 
Examples of bound entangled states are studied in the
literature; see, for example, References~\cite{PhysRevLett.80.5239,%
  PhysRevLett.113.050404,HORODECKI1997333,%
  UPBBECon,BEcon2,Kay2011,Kye2015,PhysRevA.94.020302,BEUCPB}.
They are often given as special constructions of states in some particular
parameter families.
A more general way of constructing bound entangled states which allows the
generation of a random sample has also been presented in
Reference~\cite{PhysRevA.97.032319}, but it only works for bipartite systems
with equal dimensions. 

\begin{figure}[!t]
  \centerline{\includegraphics[viewport=118 500 474 755,clip]%
    {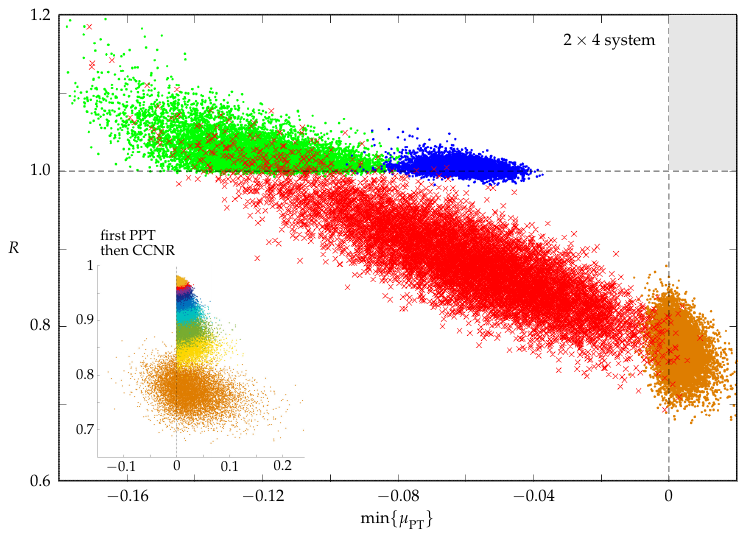}}
  \caption{\label{fig:BE2}%
    Search for bound entangled states with SCMC for the $2\times4$ system.
    The initial
  reference sample has $10^4$ states (red crosses) uniformly distributed with
  respect to the Hilbert--Schmidt distance shown here on an $R$ vs.\
  $\min\{\mu^{\ }_{\mathrm{PT}}\}$ plot.
  The states after SCMC are indicated by the blue dots.
  There are no blue dots in the first quadrant (gray) where the 
  \eqref[criteria]{BEcriteria} are obeyed. Enforcing only the PPT criterion
  results in the orange sample; enforcing only the CCNR criterion
  results in the green sample.
  The inset shows the orange sample and, in different colors, further samples
  obtained by SCMC steps toward enforcing the CCNR criterion.}
\end{figure}

 \begin{specialtable}[!b]\centering
  \caption{\label{tbl:BE}%
         Parameters used for the SCMC sampling %
         (${\color{red}\times\color{black}\to\color{blue}\bullet}$) %
         reported in Figures~\ref{fig:BE1} and \ref{fig:BE2}:
         number $N_{\tau}$ of intermediate distributions, %
         tolerance $a_e$ for the entanglement constraint, %
         tolerance $a_p$ for the PPT constraint, %
         yield of bound entangled states, and CPU time %
         per distribution step.}
\begin{tabular}{crrrcc}
\toprule
  \textbf{system}&\multicolumn{1}{c}{$\boldsymbol{N_{\tau}}$}
  &\multicolumn{1}{c}{$\boldsymbol{a_e}$}
  &\multicolumn{1}{c}{$\boldsymbol{a_p}$}
  &\textbf{yield}&\textbf{time}\\ \midrule
        $3\times3$ &    20 & 1,000 &  10,000 & 85\% & 11s\\
        $3\times4$ & 2,000 & 1,000 &  10,000 & 70\% & 14s\\
        $3\times5$ & 2,000 & 3,000 &  20,000 & 22\% & 15s\\
        $4\times4$ & 2,000 & 3,000 & 500,000 & 40\% & 22s\\
        $2\times4$ &    20 &   200 &     300 &  0\% & 7s\\
\bottomrule
\end{tabular}
\end{specialtable}

The SCMC samplers are well suited for generating random samples of PPT
bound entangled states with no restrictions on the states' parameter family or
on their dimensionality.
They can be generated by imposing the two criteria in
\eqref{BEcriteria} in the same way as one imposes the physicality
constraint. The two indicator functions are in the form of
\eqref{constrgen} with $\kappa_1(\rho)=\min\{\mu^{\ }_{\mathrm{PT}}\}$
and $\kappa_2(\rho)=R-1$, respectively.
The initial reference sample of dimension $d$ can be drawn directly from a
uniform distribution of physical states using the Wishart distribution
$W_d^{(\mathrm{Q})}(d,\mathbf{1}_d)$, that is, they are generated from
$d\times d$ normally distributed complex matrices with mean zero and
covariance matrix $\mathbf{1}_d\otimes\mathbf{1}_d$ (see
Reference~\cite{Han2021Wishart} for more details).
Since the reference samples are physical initially, and we do not want
to have unphysical sample points as the samples are processed,
we impose the hard physicality constraints during the Markov chain steps
to ensure that the random walks do not move a quantum state beyond the
physical boundary.
Examples illustrating the generation of random samples of bound entangled
states are shown in \figref{BE1}, where the states in the initial
reference sample are represented by red crosses and the sample states after
SCMC and before the final rejection sampling are represented by blue dots.
As is clearly visible from the plots, our SCMC sampler successfully moved the
states toward the region around ${R=1}$ and
${\mathrm{min}\{\mu^{\ }_{\mathrm{PT}}\}=0}$ and produced bound entangled
states.
Table \ref{tbl:BE} lists the parameters used for the SCMC sampling together
with the yield of bound entangled states and the CPU time consumed in each
distribution step.

As discussed in the previous section, the efficiency of SCMC depends on a
number of parameters, including the tolerance $a_p$ for the PPT constraint,
the tolerance $a_e$ for the CCNR entanglement constraint, and the number
$N_\tau$ of the distribution steps.
Rejection sampling is applied to impose the complete set of hard constraints
at the end of the algorithm, thus a good set of the parameters should result
in a high yield of accepted sample points within reasonable CPU time.
For the bipartite qutrit system of dimension ${d=3\times3}$, a $99\%$ yield of
bound entangled states was obtained by setting ${N_\tau= 300}$ and
${a_e=a_p=5\times10^4}$.
On our standard desktop, it took a few minutes to find thousands of
two-qutrit bound entangled states; CPU parallelization is possible and will
speed up the computation.
For comparison, we conducted independent sampling by drawing $10^{10}$ states
from a uniform distribution of bipartite qutrit states.
Only $24$ out of the $10^{10}$ states
obeyed the criteria of \eqref{BEcriteria}, and this process
took days even with CPU parallelization. 
Such a search for higher-dimensional systems ($3\times4$, $3\times5$, or
$4\times4$, say) is even more difficult.  
This evidently shows that the improvement provided by the SCMC algorithm is
tremendous. 

In Figures~\ref{fig:BE1} and \ref{fig:BE2}, we show scatter plots of the
$\min\{\mu^{\ }_{\mathrm{PT}}\}$ and $R$ values of samples obtained by SCMC for
five systems with Hilbert-space dimensions between eight and sixteen (state
space dimensions between 63 and 255). 
No exhaustive trials were done, and the performance can certainly be improved
upon as we did not optimize the parameters. 
Since we are imposing the constraints through the two inequalities in
\eqref{BEcriteria}, the states produced were clustered in the small
corner near ${R=1}$ and ${\min\{\mu^{\ }_{\mathrm{PT}}\}=0}$ for the four
systems in \figref{BE1}, \emph{but not}
for the $2\times4$ system in \figref{BE2}.
For the cases of \figref{BE1},
to explore other parts of the space for larger values of $R$ and 
$\min\{\mu^{\ }_{\mathrm{PT}}\}$, we filtered the sample further through MCMC
iterations with different kernels and acceptance criteria, such as ``accept
only if $R$ is increased;'' the samples obtained are shown in the insets of
\figref{BE1}.  
This way of producing random samples of bound entangled states, without
relying on special constructions of parameter families of states, will be very
useful when studying general properties of bound entanglement; such
investigations are, however, not the objective of this work.

The $2\times4$ system of \figref{BE2} is particular---we could not find any
states that 
satisfy both the PPT and the CCNR criterion \eqref[in]{BEcriteria}.
When either one of the criteria is enforced, SCMC yields samples with 
${\mathrm{min}\{\mu^{\ }_{\mathrm{PT}}\}\geq0}$ or ${R>1}$, respectively.
As the $2\times4$ plot in \figref{BE2} shows, first enforcing the PPT
criterion, followed by SCMC steps toward enforcing the CCNR criterion,
produces samples with $R$ values that are sequentially closer to the ${R=1}$
threshold without, however, crossing it.
We conjecture that $2\times4$ systems do not have bound entangled states that
obey the sufficient double criterion \eqref[of]{BEcriteria}.
Indeed, the families of bound entangled $2\times4$ states constructed by
the authors of \cite{HORODECKI1997333,Kay2011,Kye2015}
are PPT with ${R\leq1}$;
their values are on the straight lines with endpoints
${\bigl(\mathrm{min}\{\mu^{\ }_{\mathrm{PT}}\},R)=(0,1)}$ and $(0,0.7866)$,
$(0,0.8727)$ and $(0.0270,0.7572)$, or $(0,1)$ and $(0,0.8536)$, respectively.

These examples illustrate that the \eqref[criteria]{BEcriteria} are
not necessary---the pair is sufficient.
It is possible, perhaps likely, that the final SCMC samples (blue dots in
Figures~\ref{fig:BE1} and \ref{fig:BE2}) contain bound entangled states
outside the first quadrant, but the yields reported in Table~\ref{tbl:BE}
refer solely to the states that meet the double criterion.

\subsection{Desired target distribution}\label{sec:Desired}
Random samples from a desired distribution are useful in various contexts in
quantum information science such as studying properties of a quantum system,
parameter optimization, or model testing.  
Owing to the notorious curse of dimensionality and/or quantum constraints,
sampling methods that work well for systems of low dimension fail to work in
practice as the dimension gets larger.
(The curse refers to the dimensionality of the parameter space,
  not of the Hilbert space.)
For example, the generation of a target sample from a uniform reference
distribution of states through rejection sampling fails to work for
three-qubit systems as the acceptance rate is extremely low;  
one can increase the acceptance rate by orders of magnitude by replacing the
uniform reference sample with an appropriate Wishart distribution of states,
nevertheless, the efficiency still suffers from an exponential decay with
respect to dimensionality~\cite{Han2021Wishart}.
On the other hand, Markov chain algorithms suffer from instability and low
efficiency issues as the dimension of the quantum system gets larger, in
addition to their unavoidable sample correlation~\cite{Shang2015, Seah2015}.   
In this section, we demonstrate that SCMC can be used for generating samples
of a desired target distribution in high dimensions, impractical otherwise,
and this algorithm does not suffer from the curse of dimensionality. 
	
To be specific, we consider the generic situation where the desired
distribution is of the Dirichlet form 
\begin{equation}
  f(\rho) \propto \prod_k \mathrm{tr}\bigl(\Pi_k\rho\bigr)^{ \alpha_k}
  = \prod_k p_k^{ \alpha_k}, \quad \rho>0\,, \label{eq:Dirichlet}
\end{equation}
which is central to quantum state estimation with a conjugate prior.
The probability operators $\Pi_k$ have the usual properties, namely
$\Pi_k\geq0$ and $\sum_k\Pi_k=\mathbf{1}$, and the set
${\boldsymbol{\Pi}=\{\Pi_k\}}$
is an informationally complete measurement so that there is a one-to-one
correspondence between the state $\rho$ and the probabilities
${\mathbf{p}=\{p_k\}=\mathrm{tr}\bigl(\boldsymbol{\Pi}\rho\bigr)}$.  
Note that a sample drawn from the Dirichlet distribution by one of the
standard efficient algorithms, which draw from the probability simplex, has
many unphysical entries (${p_k>0}$ while ${\rho\not\geq0}$) and simply
discarding them would results in a very poor yield.
The SCMC procedure pushes most of the initially unphysical sample points into
the physical state space.
Similar remarks apply to samples drawn from a Dirichlet distribution
centered at the state for which $f(\rho)$ is largest, which is often a
rank-deficient state. 
Note also that the Dirichlet distribution has a single very narrow peak when
${A=\sum_k\alpha_k}$ is large, as is the typical situation in quantum state
estimation scenarios.

To apply SCMC, we first need a set of initial sample points that can be easily
generated.
Conventional MC methods often use samples generated in the probability
simplex, such as the Dirichlet distribution that do not respect the positivity
constraints, but the rate of physical samples decreases exponentially with
respect to the dimension of the parameter space which makes it impractical for
systems of high dimension (for example, the physical rate is less than
$10^{-10}$ for a three-qubit system~\cite{Han2021Wishart}).
Therefore, for efficient sampling, it is expedient to use an initial reference
sample that is physical.
We mainly explore two types of initial reference sample distributions --- the
uniform distribution with respect to the Hilbert--Schmidt distance and the
peaked Wishart distribution.
The former rarely resembles any feature of the target distribution but the
intermediate distributions $\{h_i(\rho)\}$ are straightforward to evaluate for
each sample point in the algorithm.
The Wishart distribution for quantum states $W^{(\mathrm{Q})}_d(n,\Sigma)$
offers more freedom by adjusting the covariance matrix $\Sigma$ in shaping the
reference distribution towards a target while conforming to the physicality
constraint.
Its probability density for a $d$-dimensional system is~\cite{Han2021Wishart}
\begin{equation}\label{eq:ws}
  g(\rho) \propto \frac{\det(\rho)^{n-d}}
                       {\mathrm{tr}\bigl(\Sigma^{-1}\rho\bigr)^{nd}}\,,
\end{equation}
where ${\rho=Z^\dagger Z/\mathrm{tr}\bigl(Z^\dagger Z\bigr)}$ and the $d\times
n$ random complex matrix $Z$, with ${n\geq d}$, is drawn from a gaussian
distribution with zero mean and covariance matrix
$\mathbf{1}_n\otimes\Sigma$; we get the uniform distribution for ${n=d}$ and
${\Sigma=\mathbf{1}_d}$. 
We expect faster convergence using the Wishart distribution as it is more
similar to the target distribution.
However, the apparent downside is that the numerical evaluation of its initial
and resulting intermediate distributions takes longer.  

To test its performance, we run the SCMC algorithm for different numbers of
intermediate distributions $N_\tau$ and different types of initial reference
distributions. 
We assess the quality of the sample by a variant of the method described in
\cite[Sec.~5]{Han2021Wishart}, for which we introduce the nested
$\lambda$-regions with ${f(\rho)\geq\lambda F}$ where
${F=\max_{\rho}\bigl\{f(\rho)\bigr\}}$ is the peak value of $f(\rho)$, and
${0\leq\lambda\leq1}$. 
We have the full physical state space for $\lambda=0$ and only the peak
location for $\lambda=1$.
The fraction of the target distribution contained in a $\lambda$-region is its
\emph{content}\footnote{%
  In the context of Bayesian estimation, $c^{\ }_{\lambda}$
  is the credibility of the region, if $f(\rho)$ is the posterior distribution.}
$c^{\ }_{\lambda}$ 
\begin{equation}
  \label{eq:content}
  c_{\lambda}^{\ }=\int(\D\rho)\,f(\rho)\eta\bigl(f(\rho)-\lambda F\bigr)\,,
\end{equation}
where $c_{\lambda=0}^{\ }=1$ reflects the normalization of $f(\rho)$ to unit
integral.
By counting how many sample points are inside a $\lambda$-region, we obtain an
estimate for the respective $c^{\ }_{\lambda}$ value.

\begin{figure}[!t]
  \centerline{\includegraphics[viewport=118 250 478 750,clip]%
  {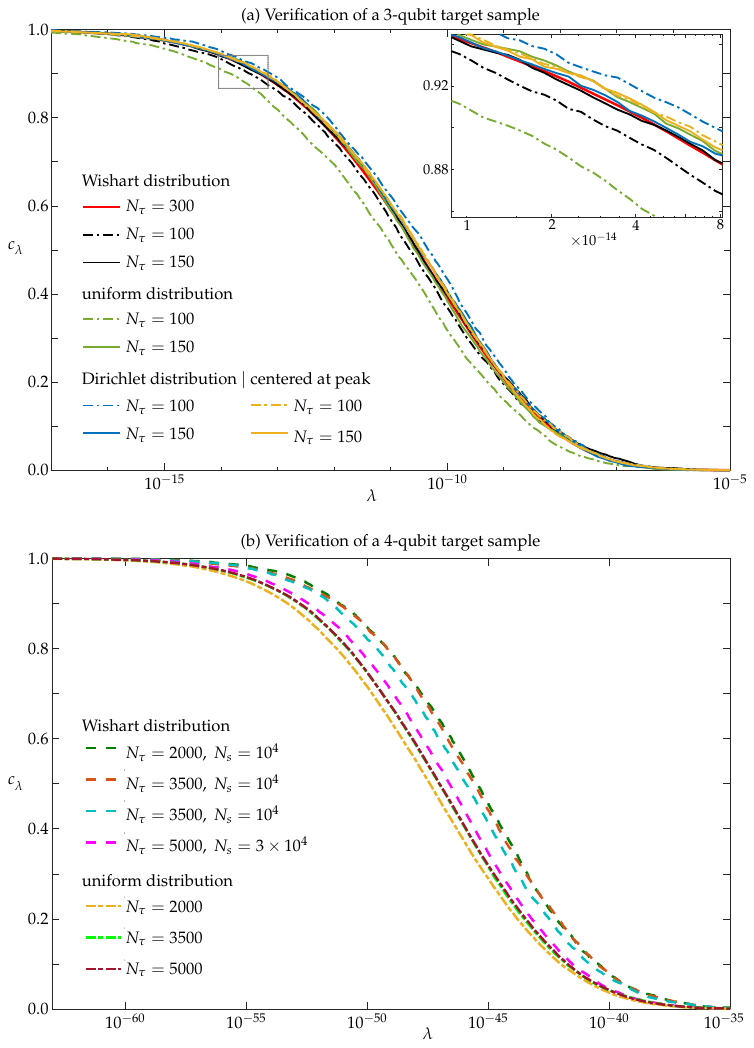}}%
\caption{\label{fig:34qbbkref}%
  The content $c_\lambda$ evaluated for samples generated for
  (a) a three-qubit target distribution and
  (b) a four-qubit target distribution.
  The target distributions are given by ${A=3000}$ randomly
  generated detection events for product tetrahedron
  measurements~\cite{34qubitdata}.
  The samples are obtained using SCMC with $N_s$ initial reference points
  drawn from the Wishart distribution, the uniform distribution, the Dirichlet
  distribution, or the Dirichlet distribution centered at the peak of $f(\rho)$.
  The SCMC algorithm is run for different numbers of intermediate
  distributions $N_\tau$.
  The inset in (a) is a blow-up of the marked rectangular area.} 
\end{figure}

We expect that this estimate is better (i) when more intermediate steps
$N_{\tau}$ are used in the SCMC algorithm, and (ii) when the sample size
$N_s$ is larger.
Both expectations are confirmed by the data presented in
\figref{34qbbkref} for two examples, one for a three-qubit system and
one for a four-qubit system.
The two $f(\rho)$s are posterior distributions (for a uniform prior) for
${A=3000}$ randomly generated measurement clicks of product tetrahedron
measurements~\cite{34qubitdata}.  

With this large $A$ value, the distributions are squeezed into a tiny region
in the immediate vicinity of the peak location of $f(\rho)$ at or near the
boundary of the state space, which makes them impractical for sampling with
conventional methods.  
In the SCMC algorithm, besides imposing the sequence of intermediate
distributions to approach the desired one gradually from the reference
distribution, the physicality constraints also have to be imposed either
strictly or gradually.
When the initial sample points are guaranteed to be physical, such as the ones
uniformly drawn from the state space, the hard physicality constraint is
directly imposed during the MCMC iterations. 
Otherwise, when a portion of the initial sample distribution is unphysical, as
in the case of the linearly shifted Wishart distribution~\cite{Han2021Wishart}
or the Dirichlet distributions, we impose the soft physicality constraint
gradually along with the distribution steps.\footnote{%
  When the constraint is strictly imposed throughout, one could make use of a
Cholesky decomposition to check for a semi-positive-definite and hermitian
matrix. 
Then, the positivity  of the smallest eigenvalue of the matrix corresponding
to $\rho$ is used to impose the soft constraint via the indicator function.}

In the three-qubit example shown in \figref{34qbbkref}(a), we find it
sufficient to have ${N_\tau=300}$ intermediate steps when the initial sample is
drawn from an appropriate Wishart distribution, as the content
barely changes by increasing $N_\tau$ further.
When ${N_\tau=100}$, the Dirichlet distribution centered at the peak of
$f(\rho)$ performs the best and the uniform distribution performs the
poorest.
Their difference becomes less and less noticeable with increasing number of
distribution steps.
For example, no significant difference of their performance is shown when
${N_\tau=150}$.
Thus, it is evident that the SCMC algorithm tolerates flexibility in the
choice of the initial reference distribution, although a reference
distribution that better resembles the target can offer faster convergence. 

\figref{34qbbkref}(b) shows the content of the four-qubit samples
generated starting from either a Wishart distribution or a uniform
distribution.  
The Dirichlet distribution is impractical to use here because the random walk
into the physical space---a tiny subspace within the 255-dimensional
probability simplex---is extremely inefficient.
For the same number of sample
points, the Wishart distribution might appear to provide faster convergence
than the uniform distribution in terms of $N_\tau$.
However, the overall performance using the uniform distribution is practically
better.
This is because the computation per sample point is about three to four
times faster when using the uniform distribution than for the Wishart
distribution. 
A larger sample size helps not only to reduce the statistical error in
evaluating sample average quantities like the content, it also
makes the random walk, which is set to propagate along the direction given by
the covariance matrix of the current sample, more efficient.\footnote{%
  The covariance matrix is more accurately estimated when the sample is larger.}
As a result, the time cost for using the Wishart distribution with
$\{N_s=3\times10^4, N_\tau=3500\}$ is about the same as using the uniform
distribution with $\{N_s=10^5, N_\tau=3500\}$ but the latter converges better,
as is visible in \figref{34qbbkref}(b). 

The computational time roughly scales as $\mathcal{O}(N_\tau\sqrt{N_s})$ with
simple vectorization of the code in Python, provided that there is sufficient
memory.
For the examples shown in \figref{34qbbkref}, the sampling of $10^5$
three-qubit states and ${N_\tau=200}$ distribution steps took about
$2\times10^4$ seconds using the Wishart distribution on a regular desktop with
no CPU-parallelization, and $7\times10^3$ seconds using the uniform
distribution.
The sampling of $10^5$ four-qubit states and ${N_\tau=3500}$ distribution steps
using the uniform distribution, which showed good convergence, took about
$2.2\times10^5$ seconds ($\sim$ 60 hours, which is only about 30 times longer
than sampling three-qubit states).  
We also ran the sampling algorithm for one-qubit and two-qubit states with
reliable sample verification.
Using the uniform distribution, the sampling of $10^5$ one-qubit states and
${N_\tau=10}$ takes about 260 seconds and the sampling of $10^5$ two-qubit
states and ${N_\tau=75}$ takes about $1.6\times10^3$
seconds~\cite{34qubitdata}. 
Similar scaling of running time was seen in other distributions we sampled.
All sampling is done on a standard desktop with \mbox{8\,GB RAM}.

In summary, for the SCMC sampling from the states spaces for one to four
qubits, which have  $(3, 15, 63, 255)$ parameters, the number of required
distribution steps is $(10, 75, 200, 3500)$
$\simeq\bigl(3^{2.1}, 15^{1.6}, 63^{1.3}, 255^{1.5}\bigr)$
and the computational time is roughly $\bigl(3^{1.3}, 15^{1.2}, 63^{1.2},
255^{1.5}\bigr)$ minutes.
The number of distribution steps required for convergence increases as the
dimensionality increases, leading to longer computational time, and additional
CPU cost is incurred by the multiplication of ever larger matrices. 
For the scaling of the running time, we observe exponents that are
approximately independent of the dimension, and this suggests that the SCMC
algorithm does not suffer an exponential increase in computation time with
respect to the dimension as other sampling methods do.
Put differently, the SCMC sampler appears to remain computationally
efficient as the dimensionality increases, at least for dimensions up to
$255$.
Owing to limited time and computer memory, we did not sample larger quantum
systems. 
However, a powerful work station running the CPU-parallelization version of
the code should be able to not only produce a larger sample in a much shorter
period of time but also samples quantum states in higher dimensions.
There is huge potential in the SCMC algorithm which is worth exploring.

\begin{figure}[!t]
  \centerline{\includegraphics[viewport=120 278 480 758,clip]%
  {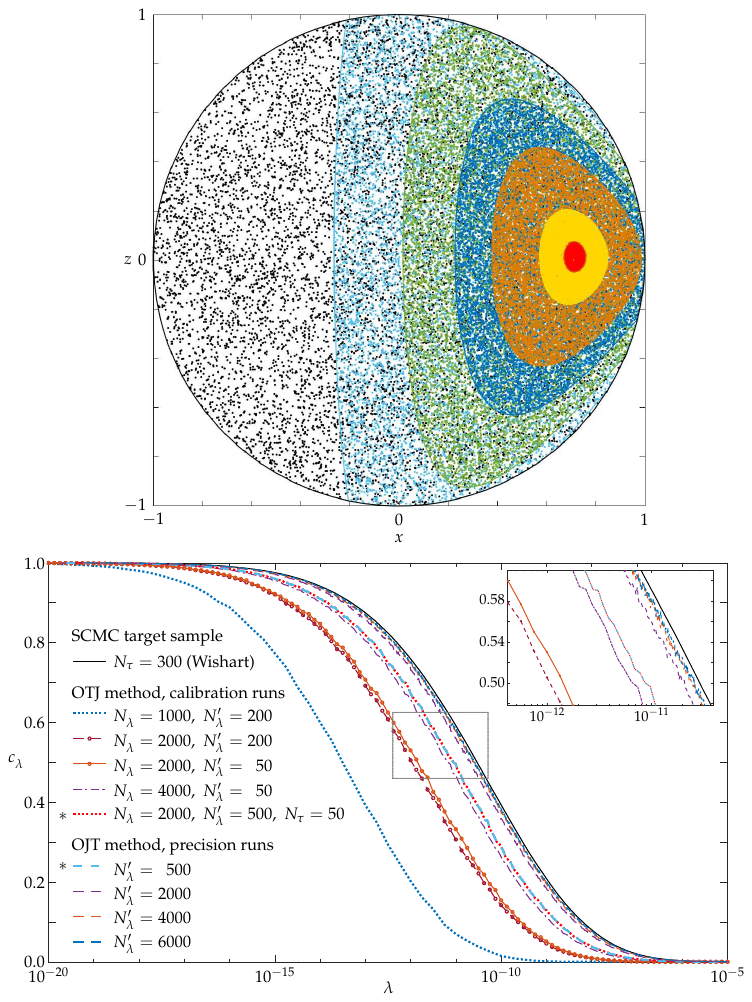}}
\caption{\label{fig:OTJ-examples}
  OTJ algorithm illustrated.
    \textbf{Top:}
    Uniform samples on the $\lambda$-regions for single-qubit states on
  the equatorial disk of the Bloch ball.
  The target distribution refers to the experimental data of a distorted trine
  measurement \cite{Len2018} with
  ${(\alpha_1,\alpha_2,\alpha_3)=(1802,315,303)}$ in \eqref{Dirichlet}.
  The scattered points of different colors mark $10^4$ states each in the
  $\lambda$-regions with with ${\log_{10}(\lambda) = -\infty}$,
  $-400$, $-200$, $-100$, $-50$, $-10$, and $-1$, respectively.
  The initial distribution (${\lambda=0}$, black points) is uniform on the disk.
  \textbf{Bottom:}
    Content $c_{\lambda}^{\ }$ evaluated for the three-qubit target distribution
    of \figref{34qbbkref}(a).
    The black curve is computed directly from a target sample with ${N_s=10^6}$
    points generated from a Wishart sample by SCMC.
    The other colored curves show results from several runs of the
    OTJ algorithm.
    The inset is a blow-up of the marked rectangular area.}
\end{figure}

\subsection{The Oh--Teo--Jeong method for benchmarking}
The volume element $(\D\rho)$ in \eqref{content} is also the probability
element of the uniform distribution, and
\begin{equation}\label{eq:size}
    s_{\lambda}^{\ }=\int(\D\rho)\,\eta\bigl(f(\rho)-\lambda F\bigr)
\end{equation}
is the \emph{size} of the $\lambda$-region, normalized such that
${s_{\lambda=0}^{\ }=1}$. 
The relation \cite{Shang2013}
\begin{equation}\label{eq:s2c}
  c^{\ }_{\lambda}=\frac{\lambda s^{\ }_{\lambda}
    +\int_{\lambda}^1\D\lambda'\, s^{\ }_{\lambda'}}
  {\int_0^1\D\lambda'\, s^{\ }_{\lambda'}}
\end{equation}
can be used to obtain an estimate of $c_{\lambda}^{\ }$ from an estimate of
$s_{\lambda}^{\ }$.
Owing to the very narrow peak of $f(\rho)$, however, we cannot estimate
$s_{\lambda}^{\ }$ by counting how many points of a uniform sample are in the
$\lambda$-region, there would be far too few sample points in the vicinity of
the peak.

The OTJ method introduced in \cite{PhysRevLett.123.040602,PhysRevA.100.012345}
offers a clever solution to this problem based on their
region-average computation lemma.
It provides $s_{\lambda}^{\ }$ by an integration,
\begin{equation}\label{eq:RAC}
  s_{\lambda}^{\ }=s_{\lambda_0}^{\ }
  \frac{g^{\ }_{\lambda_0}}{g^{\ }_{\lambda}}
  \exp\Biggl(-\int\limits_{\lambda_0}^{\lambda}
                \frac{\D\lambda'}{\lambda'g^{\ }_{\lambda'}}\Biggr)\,, 
\end{equation}
where $g^{\ }_{\lambda}$ is the average value of
$\ln\bigl(f(\rho)/(\lambda F)\bigr)$ over the $\lambda$-region with respect to
the uniform distribution.
Although $f(\rho)$  is very narrowly peaked, $\ln\bigl(f(\rho)\bigr)$ is not,
and $g^{\ }_{\lambda}$ can be estimated accurately by a sufficiently large
sample drawn from the uniform distribution over the $\lambda$-region, not over
the whole quantum state space.
Accordingly, an implementation of the OTJ method requires an accurately known
reference value $s_{\lambda_0}^{\ }$ and a reliable procedure for generating a
uniform sample for each $\lambda$-region.
Both ingredients had problems in the implementation reported in
\cite{PhysRevLett.123.040602,PhysRevA.100.012345}.

Here we implement the OTJ method by (i) finding $s_{\lambda_0}^{\ }$ from a
large uniform sample for $\lambda_0$ small enough that, at least, a few
percent of the sample points are in the $\lambda_0$-region;
and (ii) generating large uniform
samples from successive $\lambda$-regions by an SCMC algorithm that gradually
imposes the constraint ${f(\rho)\geq\lambda F}$.
For the computation of the $\lambda'$ integral in \eqref{RAC} we
discretize linearly in $\ln(\lambda')$. 

\figref{OTJ-examples} illustrates aspects of the OTJ algorithm.
At the top we see that the $\lambda$-regions are not of ellipsoidal
shape\footnote{%
  Whereas the ``accelerated hit-and-run'' algorithm used in
\cite{PhysRevLett.123.040602,PhysRevA.100.012345} assumes that the
$\lambda$-regions have ellipsoidal shape, which is only justified for
${\lambda\lesssim1}$, no such assumptions enter the SCMC algorithm.}
and that one needs truly tiny positive $\lambda$ values to enclose
a sizable fraction of a uniform sample on the whole state space.
The situation is that of a three-outcome measurement in which ${A=2420}$
events were observed~\cite[p.224]{Nott2021}.
While this is not a lot of data for a one-qubit measurement, the Bayesian
posterior (our target function here) is already peaked extremely narrowly. 

At the bottom in \figref{OTJ-examples}, we show $c^{\ }_{\lambda}$ for a
three-qubit target distribution obtained from sampling directly from the
distribution by SCMC (black curve) and by different runs of the OTJ
algorithm (colored curves).
\figref{OTJ-examples} (bottom) confirms that the SCMC sampling yields
the correct $c^{\ }_{\lambda}$ values, as the sequence of OTJ curves
converges toward the SCMC curve. 
This convergence is from below, which tells us that the OTJ estimates of
$c^{\ }_{\lambda}$ have a negative bias.

All SCMC runs of the OTJ method proceed from ${\lambda=1}$ when all sample
points are at the peak location. Then by reducing $\lambda$ in $N_{\lambda}$
steps, we obtain uniform samples on the respective $\lambda$-regions.
The ``calibration runs'' then use ${s^{\ }_{\lambda_0}=0.024}$ for
${\lambda_0=10^{-250}}$, which we get from a large uniform sample on the whole
state space, in \eqref{RAC} and the resulting $s^{\ }_{\lambda}$ in
\eqref{s2c}. 
The ``precision runs'' use $\lambda_1=10^{-25}$ in the stead of $\lambda_0$ in
\eqref{RAC}, with ${s^{\ }_{\lambda_1}\simeq10^{-24}}$ obtained from
the calibration runs, and $N'_{\lambda}$ intermediate $\lambda$ values.
The graphs for the final calibration run and the first precision run, marked
by $\ast$ in \figref{OTJ-examples} (bottom), are indistinguishable.

Regarding the computational effort, the comparison of the SCMC and the OTJ
algorithms speaks strongly in favor of SCMC.
The computation time scales roughly as
${\mathcal{O}\bigl((N_\lambda+N'_{\lambda}) N_\tau\sqrt{N_s}\bigr)}$.
Therefore, the final precision run takes about
$(2000+6000)50$ $\times\sqrt{10^3}/(200\sqrt{10^6})\simeq10^2$ times longer
than the direct production of a target sample by SCMC to evaluate
$c^{\ }_\lambda$, and the latter has higher accuracy.  

In summary, we can produce samples uniformly distributed in $\lambda$-regions
by an SCMC implementation of the OTJ method, and so evaluate the content
$c^{\ }_{\lambda}$ as a function of $\lambda$.
Our numerical results not only confirm that the OTJ method is viable, it also
demonstrates that, for a sharply peaked target distribution, it is more
efficient to evaluate the content from the target sample produced directly by
SCMC than using the more involved OTJ method.
Further, we observe that the content estimate obtained by the OTJ
method has a negative bias.

\section{Conclusion}
In conclusion, we demonstrate the reliability and efficiency of the SCMC sampler
in sampling quantum states through three explicit examples.%
\footnote{Yet another application of SCMC is reported in
  Reference \cite{Hiesmayr+3:23}.}
In the first example, we produce samples of bound entangled states
for dimensions $3\times3$, $3\times4$, $3\times5$, and $4\times4$.
For some of these cases, no such samples were available before, and our method
can also provide bound entangled states for higher dimensions.
For example, we obtain nearly ten thousand two-qutrit bound entangled
states in a few minutes by SCMC sampling, whereas several days of independence
sampling yield $24$ such states from $10^{10}$ candidates.
We also observe that the $2\times4$ system is particular, as we could
not find bound entangled states that obey the PPT-CCNR double criterion.
Based on the numerical evidence, we conjecture that such states do not exist.

In the second example, we use the SCMC method for sampling quantum states in
accordance with a given target distribution three-qubit and four-qubit
systems.
We find that this algorithm is more efficient than others, and it appears to
remain efficient even as the dimension grows.

The third example is a SCMC implementation of the OTJ method for producing
uniform samples on regions bounded by values of the target function (the
$\lambda$-regions); the shortcomings of the original implementation are avoided.
We find that, for the purpose of estimating the content of the
$\lambda$-regions, direct SCMC sampling from the target distribution is much
more efficient, and the values obtained by the OTJ method have a negative bias.

The SCMC sampler can be applied to many other sampling problems of quantum
systems as long as the constraints can be described by inequalities.
For example, one could produce samples of states that violate an inequality of
the Bell kind.
Multiple constraints can be efficiently applied in parallel.
Moreover, due to the well-known channel-state duality, SCMC sampling can also
be used to sample channels.
We invite the readers to apply the method to their specific sampling problems.

\vspace{6pt} 

\acknowledgments{%
  We would like to thank David Nott for his insightful advice on SMC.
  \mbox{J. S.} acknowledges support by the National Natural Science Foundation
  of China (Grant No.~11805010) and the Beijing Institute of Technology
  Research Fund Program for Young Scholars.
  \mbox{B.-G. E.} is extremely grateful for the long-standing support from the
  Centre for Quantum Technologies, where his share of the work was done.
  The Centre for Quantum Technologies is a Research Centre of Excellence
  funded by the Ministry of Education and the National Research Foundation of
  Singapore.
}

\clearpage%
\abbreviations{Abbreviations}%
{The following abbreviations are used in this manuscript:\\
\begin{tabular}{@{}ll@{\qquad}ll@{}}
  CCNR & Computational Cross Norm and Realignment &
  CPU  & Central Processing Unit \\
  ESS  & Effective Sample Size &
  GB   & Giga Byte\\
  HMC  & Hamiltonian Monte Carlo &
  PPT  & Positive Partial Transpose \\
  MC   & Monte Carlo &
  MCMC & Markov Chain Monte Carlo \\
  OTJ  & Oh, Teo, and Jeong &
  RAM  & Random-Access Memory \\
  SCMC & Sequentially Constrained Monte Carlo &
  SMC  & Sequential Monte Carlo 
\end{tabular}}

\appendixtitles{no} 
\appendixstart
\appendix

\section*{Appendix: SCMC algorithm}
Generally speaking, you implement SCMC as follows.
First, try to generate a sample that is of considerable size.
Pick the distribution step according to the scaling with the dimension $d$ of
the parameter space, i.e., $\propto d^{1.5}$
(see the last paragraph in \secref{Desired}).
Then, construct the intermediate distributions by the recipe of \eqref{hgen}.
Parameterize the inequalities imposed on the target sample in accordance with
\eqref{constrgen}.
By trial and error, adjust the Markov chain step size to reach an acceptance
rate around $23.4\%$.
Perhaps check the convergence by increasing the number of intermediate
distributions; once convergence is ensured, optimize for efficiency by
reducing the number of intermediate distributions.

The general algorithm takes in a set of $N_s$ initial sample points with
distribution $g(\mathbf{x})$ and finds a discretized sequence of density
distributions $\{h_i(\mathbf{x})\}$ that ``smoothly'' bridges between
$g(\mathbf{x})$ and $f(\mathbf{x})$.
$\{h_i(\mathbf{x})\}$ takes into account both the change of the probability
density distribution and the incorporation of the soft constraints given by
\eqref[Equations]{hgen} \eqref[to]{hgen2}. 

In general, there can be $N_{\mathrm{MC}}$ number of MCMC iterations between
each distribution step and importance resampling is performed whenever the
effective sample size (ESS) drops below a chosen threshold
$N_\mathrm{thres}$.
We employ the iteration kernel in the Metropolis algorithm which results in
sample points with updated weights for the $k$th iteration
${w_i^{(k)}\propto\bigl[h_i(\mathbf{x}_i^{(k)})\bigr]%
\big/\bigl[h_{i-1}(\mathbf{x}_i^{(k)})\bigr]}$. 
The ESS, that is
${\mathrm{ESS}^{-1}=\sum_{k=1}^{N_s}\bigl[w_i^{(k)}\bigr]^2}$, is a measure of
sample degeneracy~\cite{Liu1998}, i.e., a small ESS indicates only a small
portion of the sample points are filtered into the next step.
This degeneracy can be much suppressed in SMC with finer distribution steps as
compared with MCMC.
In addition, the SMC algorithms can be conveniently implemented with parallel
computation for speed-up.
The general outline of our SCMC algorithm is as follows:

\vspace{8pt plus 2pt minus 2pt}

{\small%
  \noindent\underline{\makebox[\columnwidth][c]{\bfseries%
      Sequentially constrained Monte Carlo algorithm}}
\begin{algorithmic}[1]
 \State Generate an initial sample $\{\rho^{(1:N_s)} \}$ with distribution
 $g(\rho)$  
 \State Assign weight $w^{(k)}_0 = 1/N_s$ for $k=1:N_s$
 \For {$i = 1:N_{\tau}$}
 \State Evaluate $h_i(\rho^{(1:N_s)})$ with indicators 
        $I_{\kappa,i}(\rho^{(1:N_s)})$
 \For {$j=1:N_{\mathrm{MC}}$} 
 \If {$j = 1$ } 
 \State Update $w_i^{(k)} = w_{i-1}^{(k)}
                \frac{h_i(\rho^{(k)})}{h_{i-1}(\rho^{(k)})}$
 for $k=1:N_s$
 \State Normalize $w_i^{(k)}$
 \State Calculate
    $\mathrm{ESS}=\Bigl[\sum_{k=1}^{N_s}\bigl(w_i^{(k)}\bigr)^2\Bigr]^{-1}$
 \If {$\mathrm{ESS} < N_{\mathrm{thres}}$}
 \State Importance resampling according to $w_i^{(1:N_s)}$
 \State Reset $w^{(1:N_s)}_i = 1/N_s$
 \EndIf
 \EndIf
 \State Propagate $\{\rho^{(1:N_s)}\}$ with MCMC transition kernels
 \State Accept/reject the new points
 \EndFor
 \EndFor
\end{algorithmic}\vspace*{-1.0\baselineskip}\noindent\hrulefill}

\vspace{8pt plus 2pt minus 2pt}

\begin{figure}[!t]
  \centerline{\includegraphics[viewport=116 527 474 755,clip]%
    {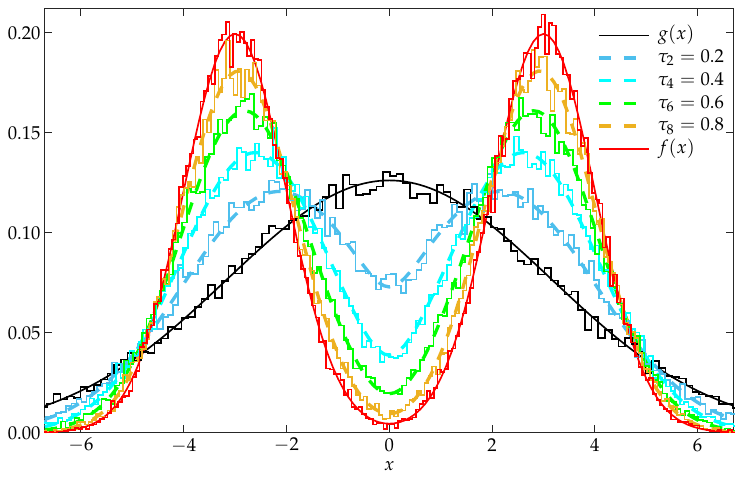}}
  \caption{\label{fig:camel}%
    A simple illustration of SCMC in 1D.
    A single-peak reference sample is used to sample a target distribution with
    two peaks.
    The intermediate distributions are given by \eqref{hgen} with
    ${N_\tau=10}$.
    Both the exact and the numerically estimated distribution for $10^4$ sample
    points are shown.}
\end{figure}

\begin{figure}[!b]
  \centerline{\includegraphics[viewport=125 585 470 752,clip]%
    {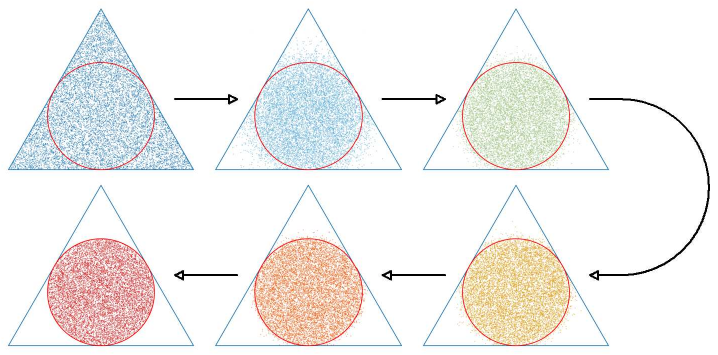}}
\caption{\label{fig:TrineDemo}%
  The physicality constraint is gradually imposed through
  ${N_\tau=10}$ intermediate distribution steps.
  The triangle represents the boundary of the probability simplex, while the
  circle represents the boundary of the physical region.
  The distributions are shown for the initial sample (uniform on the simplex,
  top left, ${\tau_0=0}$), for the final sample (uniform on the unit disk,
  bottom left, ${\tau_{10}=1}$), and for four intermediate samples
  (${\tau_i=i/10}$  with ${i=2,4,6,8}$).}
\end{figure}

\figref{camel} illustrates a simple one-dimensional example of SMC. 
The target distribution $f(x)$ has two peaks and we sample it by starting with
random samples drawn from a single-peak normal distribution $g(x)$. 
A simple example that illustrates how the physicality constraint can be
effectively imposed is shown in \figref{TrineDemo}.  
The reference sample is uniform in the probability simplex of the distorted
trine measurement of Reference~\cite{Len2018} on a single-qubit system and it is
filtered through ${N_\tau=10}$ intermediate constrained uniform distributions
thereby imposing the physicality constraint by taking $\kappa(\rho)$ to be the
smallest eigenvalue of the density matrix of state $\rho$.  

The computational efficiency of the SCMC algorithms depends on a set of
parame\-ters---the number of reference sample points $N_s$, the number of
distribution steps $N_{\tau}$, the number of Markov chain iterations
$N_{\mathrm{MC}}$ that the entire sample is advanced by a kernel of choice in each
  distribution step,\footnote{For example, the most standard gaussian kernel.
  The kernel we are using here is the ``adaptive Metropolis kernel''
  \cite[Sec.~2]{Roberts+1:09}, which has its origin in
  Reference~\cite{Haario+2:01}.}  
and the scaling parameter $a$---which need to be calibrated for the
sampling problem at hand.
For most of the examples presented in this paper, we fix $N_{\mathrm{MC}}=15$
and adjust other parameters in accordance with the criterion set
out in Reference~\cite{MCstepNew}.\footnote{%
Generally, the number of Markov chain iterations, $N_\mathrm{MC}$,
is chosen according to \cite{MCstepNew}, 		
$ N_R = \bigl\lceil\log_{1-ar}(q)\bigr\rceil$,
where $N_R$ is the required number of Markov chain steps such that there is a
$(1-q)$ probability that each point is moved at least once during the Markov
chain  procedure which has an acceptance rate of $ar$.
To put in some numbers for illustration, ${N_\mathrm{MC}= 7}$ when ${q=0.01}$ 
and ${ar=0.25}$ for a typical Metropolis--Hastings Markov chain, which has a
most efficient acceptance rate of $~0.234$ \cite{MCar}, and we choose
${N_{\mathrm{MC}}=15}$ for convenience throughout the paper.}

To reduce the sample degeneracy without compromising its efficiency by too
much, we set the threshold of ESS at $N_{\mathrm{thres}}=\frac{4}{5}N_s$ and
importance resampling is executed once ESS falls below $N_{\mathrm{thres}}$.  
The sample correlation resulting from importance resampling is attended to by
the Markov chain iterations that follow.
To reduce the correlation from the final importance resampling before the SCMC
algorithm ends, we filter the system through 20 additional Markov chain
iterations.  

\end{paracol}
\begin{raggedright}
\reftitle{References}

\end{raggedright}

\begin{thebibliography}{999}

\bibitem[Hradil(1997)]{PhysRevA.55.R1561}
Hradil, Z.
\newblock Quantum-state estimation.
\newblock {\em Phys. Rev. A} {\bf 1997}, {\em 55},~1561--1564(R).
\newblock
  doi:{\changeurlcolor{black}\href{https://doi.org/10.1103/PhysRevA.55.R1561}{\detokenize{10.1103/PhysRevA.55.R1561}}}.

\bibitem[\ifmmode \check{R}\else \v{R}\fi{}eh\'a\ifmmode~\check{c}\else
  \v{c}\fi{}ek \em{et~al.}(2015)\ifmmode \check{R}\else
  \v{R}\fi{}eh\'a\ifmmode~\check{c}\else \v{c}\fi{}ek, Teo, and
  Hradil]{PhysRevA.92.012108}
\ifmmode \check{R}\else \v{R}\fi{}eh\'a\ifmmode~\check{c}\else \v{c}\fi{}ek,
  J.; Teo, Y.S.; Hradil, Z.
\newblock Determining which quantum measurement performs better for state
  estimation.
\newblock {\em Phys. Rev. A} {\bf 2015}, {\em 92},~012108.
\newblock
  doi:{\changeurlcolor{black}\href{https://doi.org/10.1103/PhysRevA.92.012108}{\detokenize{10.1103/PhysRevA.92.012108}}}.

\bibitem[Wiebe and Granade(2016)]{PhysRevLett.117.010503}
Wiebe, N.; Granade, C.
\newblock {Efficient Bayesian phase estimation}.
\newblock {\em Phys. Rev. Lett.} {\bf 2016}, {\em 117},~010503.
\newblock
  doi:{\changeurlcolor{black}\href{https://doi.org/10.1103/PhysRevLett.117.010503}{\detokenize{10.1103/PhysRevLett.117.010503}}}.

\bibitem[Horodecki \em{et~al.}(2008)Horodecki, Horodecki, Horodecki, Leung, and
  Oppenheim]{PhysRevLett.100.110502}
Horodecki, K.; Horodecki, M.; Horodecki, P.; Leung, D.; Oppenheim, J.
\newblock Unconditional Privacy over Channels which Cannot Convey Quantum
  Information.
\newblock {\em Phys. Rev. Lett.} {\bf 2008}, {\em 100},~110502.
\newblock
  doi:{\changeurlcolor{black}\href{https://doi.org/10.1103/PhysRevLett.100.110502}{\detokenize{10.1103/PhysRevLett.100.110502}}}.

\bibitem[Hamma \em{et~al.}(2012)Hamma, Santra, and
  Zanardi]{PhysRevLett.109.040502}
Hamma, A.; Santra, S.; Zanardi, P.
\newblock Quantum Entanglement in Random Physical States.
\newblock {\em Phys. Rev. Lett.} {\bf 2012}, {\em 109},~040502.
\newblock
  doi:{\changeurlcolor{black}\href{https://doi.org/10.1103/PhysRevLett.109.040502}{\detokenize{10.1103/PhysRevLett.109.040502}}}.

\bibitem[{Dupuis} \em{et~al.}(2015){Dupuis}, {Fawzi}, and {Wehner}]{6967820}
{Dupuis}, F.; {Fawzi}, O.; {Wehner}, S.
\newblock Entanglement Sampling and Applications.
\newblock {\em IEEE Trans. Inf. Theory} {\bf 2015}, {\em 61},~1093--1112.

\bibitem[Lund \em{et~al.}(2017)Lund, Bremner, and Ralph]{Lund2017}
Lund, A.P.; Bremner, M.J.; Ralph, T.C.
\newblock Quantum sampling problems, Boson Sampling and quantum supremacy.
\newblock {\em npj Quantum Info.} {\bf 2017}, {\em 3},~15.
\newblock
  doi:{\changeurlcolor{black}\href{https://doi.org/10.1038/s41534-017-0018-2}{\detokenize{10.1038/s41534-017-0018-2}}}.

\bibitem[Bouland \em{et~al.}(2019)Bouland, Fefferman, Nirkhe, and
  Vazirani]{Bouland2019}
Bouland, A.; Fefferman, B.; Nirkhe, C.; Vazirani, U.
\newblock On the complexity and verification of quantum random circuit
  sampling.
\newblock {\em Nat. Phys.} {\bf 2019}, {\em 15},~159--163.
\newblock
  doi:{\changeurlcolor{black}\href{https://doi.org/10.1038/s41567-018-0318-2}{\detokenize{10.1038/s41567-018-0318-2}}}.

\bibitem[{Arute \textit{et al.\/}}(2019)]{shortGoogleQSupremacy}
{Arute \textit{et al.\/}}, F.
\newblock Quantum supremacy using a pro\-gramm\-able superconducting processor.
\newblock {\em Nature} {\bf 2019}, {\em 574},~505--510.
\newblock
  doi:{\changeurlcolor{black}\href{https://doi.org/10.1038/s41586-019-1666-5}{\detokenize{10.1038/s41586-019-1666-5}}}.

\bibitem[Donoho(2000)]{Donoho2000}
Donoho, D.L.
\newblock High-dimensional data analysis: The curses and blessings of
  dimensionality.
\newblock  AMS Math Challenges Lecture,  2000, pp. 1--32.

\bibitem[Wootters(1990)]{Wootters1990}
Wootters, W.K.
\newblock Random quantum states.
\newblock {\em Found. Phys.} {\bf 1990}, {\em 20},~1365--1378.
\newblock
  doi:{\changeurlcolor{black}\href{https://doi.org/10.1007/BF01883491}{\detokenize{10.1007/BF01883491}}}.

\bibitem[Bengtsson and {\.{Z}}yczkowski(2006)]{Bengtsson2006}
Bengtsson, I.; {\.{Z}}yczkowski, K.
\newblock {\em Geometry of Quantum States: An Introduction to Quantum
  Entanglement}; Cambridge University Press,  2006.
\newblock
  doi:{\changeurlcolor{black}\href{https://doi.org/10.1017/CBO9780511535048}{\detokenize{10.1017/CBO9780511535048}}}.

\bibitem[Osipov \em{et~al.}(2010)Osipov, Sommers, and
  {\.{Z}}yczkowski]{AlOsipov2010}
Osipov, V.A.; Sommers, H.J.; {\.{Z}}yczkowski, K.
\newblock Random {B}ures mixed states and the distribution of their purity.
\newblock {\em J. Phys. A} {\bf 2010}, {\em 43},~055302.
\newblock
  doi:{\changeurlcolor{black}\href{https://doi.org/10.1088/1751-8113/43/5/055302}{\detokenize{10.1088/1751-8113/43/5/055302}}}.

\bibitem[{\.{Z}}yczkowski \em{et~al.}(2011){\.{Z}}yczkowski, Penson, Nechita,
  and Collins]{Zyczkowski2011}
{\.{Z}}yczkowski, K.; Penson, K.A.; Nechita, I.; Collins, B.
\newblock Generating random density matrices.
\newblock {\em J. Math. Phys.} {\bf 2011}, {\em 52},~062201.
\newblock
  doi:{\changeurlcolor{black}\href{https://doi.org/10.1063/1.3595693}{\detokenize{10.1063/1.3595693}}}.

\bibitem[Collins and Nechita(2016)]{Collins2016}
Collins, B.; Nechita, I.
\newblock Random matrix techniques in quantum information theory.
\newblock {\em J. Math. Phys.} {\bf 2016}, {\em 57},~015215.
\newblock
  doi:{\changeurlcolor{black}\href{https://doi.org/10.1063/1.4936880}{\detokenize{10.1063/1.4936880}}}.

\bibitem[Wishart(1928)]{Wishart1928}
Wishart, J.
\newblock {The generalised product moment distribution in samples from a normal
  multivariate population}.
\newblock {\em Biometrika} {\bf 1928}, {\em 20A},~32--52.
\newblock
  doi:{\changeurlcolor{black}\href{https://doi.org/10.1093/biomet/20A.1-2.32}{\detokenize{10.1093/biomet/20A.1-2.32}}}.

\bibitem[Goodman(1963{\natexlab{a}})]{Goodman1963a}
Goodman, N.R.
\newblock {Statistical analysis based on a certain multivariate complex
  gaussian distribution (An introduction)}.
\newblock {\em The Annals of Mathematical Statistics} {\bf 1963}, {\em 34},~152
  -- 177.
\newblock
  doi:{\changeurlcolor{black}\href{https://doi.org/10.1214/aoms/1177704250}{\detokenize{10.1214/aoms/1177704250}}}.

\bibitem[Goodman(1963{\natexlab{b}})]{Goodman1963b}
Goodman, N.R.
\newblock {The distribution of the determinant of a complex {W}ishart
  distributed matrix}.
\newblock {\em The Annals of Mathematical Statistics} {\bf 1963}, {\em 34},~178
  -- 180.
\newblock
  doi:{\changeurlcolor{black}\href{https://doi.org/10.1214/aoms/1177704251}{\detokenize{10.1214/aoms/1177704251}}}.

\bibitem[Han \em{et~al.}()Han, Li, Bagchi, Ng, and Eng\-lert]{Han2021Wishart}
Han, R.; Li, W.; Bagchi, S.; Ng, H.K.; Eng\-lert, B.G.
\newblock {Uncorrelated problem-specific samples of quantum states from
  zero-mean Wishart distributions}.
\newblock {\em Eprint {\normalfont\bfseries2021} arXiv:2106.08533}.
\newblock
  doi:{\changeurlcolor{black}\href{https://doi.org/10.48550/arXiv.2106.08533}{\detokenize{10.48550/arXiv.2106.08533}}}.

\bibitem[Hastings(1970)]{Hastings1970}
Hastings, W.K.
\newblock Monte {C}arlo sampling methods using {M}arkov chains and their
  applications.
\newblock {\em Biometrika} {\bf 1970}, {\em 57},~97--109.

\bibitem[Neal(2011)]{neal2012mcmc}
Neal, R.M., {MCMC using Hamiltonian dynamics}.
\newblock In {\em {Handbook of Markov Chain Monte Carlo}}; Steve~Brooks,
  Andrew~Gelman, G.L.J.; Meng, X.L., Eds.; Chapman \& Hall/CRC,  2011;
  chapter~5, pp. 113--162.

\bibitem[Shang \em{et~al.}(2015)Shang, Seah, Ng, Nott, and
  Eng\-lert]{Shang2015}
Shang, J.; Seah, Y.L.; Ng, H.K.; Nott, D.J.; Eng\-lert, B.G.
\newblock {Monte Carlo sampling from the quantum state space. I}.
\newblock {\em New J. of Phys.} {\bf 2015}, {\em 17},~043017.
\newblock
  doi:{\changeurlcolor{black}\href{https://doi.org/10.1088/1367-2630/17/4/043017}{\detokenize{10.1088/1367-2630/17/4/043017}}}.

\bibitem[Seah \em{et~al.}(2015)Seah, Shang, Ng, Nott, and Eng\-lert]{Seah2015}
Seah, Y.L.; Shang, J.; Ng, H.K.; Nott, D.J.; Eng\-lert, B.G.
\newblock {Monte Carlo sampling from the quantum state space. II}.
\newblock {\em New J. of Phys.} {\bf 2015}, {\em 17},~043018.
\newblock
  doi:{\changeurlcolor{black}\href{https://doi.org/10.1088/1367-2630/17/4/043018}{\detokenize{10.1088/1367-2630/17/4/043018}}}.

\bibitem[Golchi and Campbell(2016)]{GOLCHI201698}
Golchi, S.; Campbell, D.A.
\newblock {Sequentially constrained Monte Carlo}.
\newblock {\em Comput. Stat. Data Anal.} {\bf 2016}, {\em 97},~98--113.

\bibitem[Del~Moral \em{et~al.}(2006)Del~Moral, Doucet, and Jasra]{oSMC}
Del~Moral, P.; Doucet, A.; Jasra, A.
\newblock {Sequential Monte Carlo samplers}.
\newblock {\em J. R. Statist. Soc. B} {\bf 2006}, {\em 68},~411--436.

\bibitem[Doucet \em{et~al.}(2001)Doucet, Freitas, and Gordon]{Doucet2001}
Doucet, A.; Freitas, N.; Gordon, N., Eds.
\newblock {\em {Sequential Monte Carlo Methods in Practice}}; Springer New
  York,  2001.
\newblock
  doi:{\changeurlcolor{black}\href{https://doi.org/10.1007/978-1-4757-3437-9}{\detokenize{10.1007/978-1-4757-3437-9}}}.

\bibitem[Chopin(2002)]{Chopin2002}
Chopin, N.
\newblock {A sequential particle filter method for static models}.
\newblock {\em Biometrika} {\bf 2002}, {\em 89},~539--552.

\bibitem[Ralph \em{et~al.}(2017)Ralph, Maskell, and Jacobs]{smcParaEst1}
Ralph, J.F.; Maskell, S.; Jacobs, K.
\newblock {Multiparameter estimation along quantum trajectories with sequential
  Monte Carlo methods}.
\newblock {\em Phys. Rev. A} {\bf 2017}, {\em 96},~052306.
\newblock
  doi:{\changeurlcolor{black}\href{https://doi.org/10.1103/PhysRevA.96.052306}{\detokenize{10.1103/PhysRevA.96.052306}}}.

\bibitem[Husz\'ar and Houlsby(2012)]{PhysRevA.85.052120}
Husz\'ar, F.; Houlsby, N.M.T.
\newblock {Adaptive Bayesian quantum tomography}.
\newblock {\em Phys. Rev. A} {\bf 2012}, {\em 85},~052120.
\newblock
  doi:{\changeurlcolor{black}\href{https://doi.org/10.1103/PhysRevA.85.052120}{\detokenize{10.1103/PhysRevA.85.052120}}}.

\bibitem[Ralph \em{et~al.}(2017)Ralph, Maskell, and Jacobs]{PhysRevA.96.052306}
Ralph, J.F.; Maskell, S.; Jacobs, K.
\newblock {Multiparameter estimation along quantum trajectories with sequential
  Monte Carlo methods}.
\newblock {\em Phys. Rev. A} {\bf 2017}, {\em 96},~052306.
\newblock
  doi:{\changeurlcolor{black}\href{https://doi.org/10.1103/PhysRevA.96.052306}{\detokenize{10.1103/PhysRevA.96.052306}}}.

\bibitem[Oh \em{et~al.}(2019{\natexlab{a}})Oh, Teo, and
  Jeong]{PhysRevLett.123.040602}
Oh, C.; Teo, Y.S.; Jeong, H.
\newblock {Probing Bayesian credible regions intrinsically: A feasible error
  certification for physical systems}.
\newblock {\em Phys. Rev. Lett.} {\bf 2019}, {\em 123},~040602.
\newblock
  doi:{\changeurlcolor{black}\href{https://doi.org/10.1103/PhysRevLett.123.040602}{\detokenize{10.1103/PhysRevLett.123.040602}}}.

\bibitem[Oh \em{et~al.}(2019{\natexlab{b}})Oh, Teo, and
  Jeong]{PhysRevA.100.012345}
Oh, C.; Teo, Y.S.; Jeong, H.
\newblock {Efficient Bayesian credible-region certification for quantum-state
  tomography}.
\newblock {\em Phys. Rev. A} {\bf 2019}, {\em 100},~012345.
\newblock
  doi:{\changeurlcolor{black}\href{https://doi.org/10.1103/PhysRevA.100.012345}{\detokenize{10.1103/PhysRevA.100.012345}}}.

\bibitem[SCM()]{SCMCrepository}
See at https://github.com/feuerbutter/QSCMC.

\bibitem[Gekman and Meng(1998)]{Gelman1998}
Gekman, A.; Meng, X.L.
\newblock Simulating normalizing constants: from importance sampling to bridge
  sampling to path sampling.
\newblock {\em Statist. Sci.} {\bf 1998}, {\em 13},~163--185.

\bibitem[Neal(2001)]{Neal2001}
Neal, R.M.
\newblock Annealed importance sampling.
\newblock {\em Stat. Comput.} {\bf 2001}, {\em 11},~125--139.
\newblock
  doi:{\changeurlcolor{black}\href{https://doi.org/10.1023/A:1008923215028}{\detokenize{10.1023/A:1008923215028}}}.

\bibitem[Horodecki \em{et~al.}(1998)Horodecki, Horodecki, and
  Horodecki]{PhysRevLett.80.5239}
Horodecki, M.; Horodecki, P.; Horodecki, R.
\newblock {Mixed-state entanglement and distillation: Is there a ``bound''
  entanglement in nature?}
\newblock {\em Phys. Rev. Lett.} {\bf 1998}, {\em 80},~5239--5242.
\newblock
  doi:{\changeurlcolor{black}\href{https://doi.org/10.1103/PhysRevLett.80.5239}{\detokenize{10.1103/PhysRevLett.80.5239}}}.

\bibitem[Hiesmayr \em{et~al.}(2025)Hiesmayr, Popp, and Sutter]{Hiesmayr+2:25}
Hiesmayr, B.C.; Popp, C.; Sutter, T.C.
\newblock Bipartite bound entanglement.
\newblock {\em Int. J. Quantum Inf.} {\bf 2025}, {\em 23},~2530003.
\newblock
  doi:{\changeurlcolor{black}\href{https://doi.org/10.1142/S0219749925300037}{\detokenize{10.1142/S0219749925300037}}}.

\bibitem[Moroder \em{et~al.}(2014)Moroder, Gittsovich, Huber, and
  G\"uhne]{PhysRevLett.113.050404}
Moroder, T.; Gittsovich, O.; Huber, M.; G\"uhne, O.
\newblock {Steering bound entangled states: A counterexample to the stronger
  Peres conjecture}.
\newblock {\em Phys. Rev. Lett.} {\bf 2014}, {\em 113},~050404.
\newblock
  doi:{\changeurlcolor{black}\href{https://doi.org/10.1103/PhysRevLett.113.050404}{\detokenize{10.1103/PhysRevLett.113.050404}}}.

\bibitem[Tendick \em{et~al.}(2020)Tendick, Kampermann, and
  Bru\ss{}]{PhysRevLett.124.050401}
Tendick, L.; Kampermann, H.; Bru\ss{}, D.
\newblock {Activation of nonlocality in bound entanglement}.
\newblock {\em Phys. Rev. Lett.} {\bf 2020}, {\em 124},~050401.
\newblock
  doi:{\changeurlcolor{black}\href{https://doi.org/10.1103/PhysRevLett.124.050401}{\detokenize{10.1103/PhysRevLett.124.050401}}}.

\bibitem[Horodecki \em{et~al.}(1996)Horodecki, Horodecki, and
  Horodecki]{HORODECKI19961}
Horodecki, M.; Horodecki, P.; Horodecki, R.
\newblock Separability of mixed states: necessary and sufficient conditions.
\newblock {\em Phys. Lett. A} {\bf 1996}, {\em 223},~1--8.
\newblock
  doi:{\changeurlcolor{black}\href{https://doi.org/https://doi.org/10.1016/S0375-9601(96)00706-2}{\detokenize{https://doi.org/10.1016/S0375-9601(96)00706-2}}}.

\bibitem[Horodecki \em{et~al.}(2005)Horodecki, Horodecki, Horodecki, and
  Oppenheim]{PhysRevLett.94.200501}
Horodecki, K.; Horodecki, M.; Horodecki, P.; Oppenheim, J.
\newblock {Locking entanglement with a single qubit}.
\newblock {\em Phys. Rev. Lett.} {\bf 2005}, {\em 94},~200501.
\newblock
  doi:{\changeurlcolor{black}\href{https://doi.org/10.1103/PhysRevLett.94.200501}{\detokenize{10.1103/PhysRevLett.94.200501}}}.

\bibitem[Czekaj \em{et~al.}(2015)Czekaj, Przysi\c{e}\.{z}na, Horodecki, and
  Ho\-ro\-dec\ ki]{Czekaj2015}
Czekaj, L.; Przysi\c{e}\.{z}na, A.; Horodecki, M.; Ho\-ro\-dec\ ki, P.
\newblock {Quantum metrology: Heisenberg limit with bound entanglement}.
\newblock {\em Phys. Rev. A} {\bf 2015}, {\em 92},~062303.
\newblock
  doi:{\changeurlcolor{black}\href{https://doi.org/10.1103/PhysRevA.92.062303}{\detokenize{10.1103/PhysRevA.92.062303}}}.

\bibitem[Chen and Wu(2003)]{CCNR}
Chen, K.; Wu, L.A.
\newblock {A matrix realignment method for recognizing entanglement}.
\newblock {\em Quantum Inf. Comput.} {\bf 2003}, {\em 3},~193--202.

\bibitem[Horodecki(1997)]{HORODECKI1997333}
Horodecki, P.
\newblock Separability criterion and inseparable mix\-ed states with positive
  partial transposition.
\newblock {\em Phys. Lett. A} {\bf 1997}, {\em 232},~333--339.
\newblock
  doi:{\changeurlcolor{black}\href{https://doi.org/https://doi.org/10.1016/S0375-9601(97)00416-7}{\detokenize{https://doi.org/10.1016/S0375-9601(97)00416-7}}}.

\bibitem[Bennett \em{et~al.}(1999)Bennett, DiVincenzo, Mor, Shor, Smolin, and
  Terhal]{UPBBECon}
Bennett, C.H.; DiVincenzo, D.P.; Mor, T.; Shor, P.W.; Smolin, J.A.; Terhal,
  B.M.
\newblock {Unextendible product bases and bound entanglement}.
\newblock {\em Phys. Rev. Lett.} {\bf 1999}, {\em 82},~5385--5388.
\newblock
  doi:{\changeurlcolor{black}\href{https://doi.org/10.1103/PhysRevLett.82.5385}{\detokenize{10.1103/PhysRevLett.82.5385}}}.

\bibitem[Bru\ss{} and Peres(2000)]{BEcon2}
Bru\ss{}, D.; Peres, A.
\newblock {Construction of quantum states with bound entanglement}.
\newblock {\em Phys. Rev. A} {\bf 2000}, {\em 61},~030301.
\newblock
  doi:{\changeurlcolor{black}\href{https://doi.org/10.1103/PhysRevA.61.030301}{\detokenize{10.1103/PhysRevA.61.030301}}}.

\bibitem[Kay(2011)]{Kay2011}
Kay, A.
\newblock Optimal detection of entanglement in
  {G}reen\-ber\-ger--{H}orne--{Z}ei\-linger states.
\newblock {\em Phys. Rev. A} {\bf 2011}, {\em 83},~020303.
\newblock
  doi:{\changeurlcolor{black}\href{https://doi.org/10.1103/PhysRevA.83.020303}{\detokenize{10.1103/PhysRevA.83.020303}}}.

\bibitem[Kye(2015)]{Kye2015}
Kye, S.H.
\newblock Three-qubit entanglement witnesses with the full spanning properties.
\newblock {\em J. Phys. A: Math.\ Theor.} {\bf 2015}, {\em 48},~235303.
\newblock
  doi:{\changeurlcolor{black}\href{https://doi.org/10.1088/1751-8113/48/23/235303}{\detokenize{10.1088/1751-8113/48/23/235303}}}.

\bibitem[Sent\'{\i}s \em{et~al.}(2016)Sent\'{\i}s, Eltschka, and
  Siewert]{PhysRevA.94.020302}
Sent\'{\i}s, G.; Eltschka, C.; Siewert, J.
\newblock Quantitative bound entanglement in two-qutrit states.
\newblock {\em Phys. Rev. A} {\bf 2016}, {\em 94},~020302.
\newblock
  doi:{\changeurlcolor{black}\href{https://doi.org/10.1103/PhysRevA.94.020302}{\detokenize{10.1103/PhysRevA.94.020302}}}.

\bibitem[Halder and Sengupta(2019)]{BEUCPB}
Halder, S.; Sengupta, R.
\newblock {Construction of noisy bound entangled states and the range
  criterion}.
\newblock {\em Phys. Lett. A} {\bf 2019}, {\em 383},~2004--2010.
\newblock
  doi:{\changeurlcolor{black}\href{https://doi.org/https://doi.org/10.1016/j.physleta.2019.04.003}{\detokenize{https://doi.org/10.1016/j.physleta.2019.04.003}}}.

\bibitem[Sindici and Piani(2018)]{PhysRevA.97.032319}
Sindici, E.; Piani, M.
\newblock Simple class of bound entangled states based on the properties of the
  antisymmetric subspace.
\newblock {\em Phys. Rev. A} {\bf 2018}, {\em 97},~032319.
\newblock
  doi:{\changeurlcolor{black}\href{https://doi.org/10.1103/PhysRevA.97.032319}{\detokenize{10.1103/PhysRevA.97.032319}}}.

\bibitem[34q()]{34qubitdata}
The sequence of randomly generated measurement clicks data (${A=3000}$) for
  product tetrahedron measurements in lexico\-graphic order: the data for the
  one-qubit example are \{1135, 1086, 394, 385\}; the data for the two-qubit
  example are \{215, 34, 229, 248, 52, 244, 235, 256, 257, 225, 234, 41, 217,
  238, 44, 231\}; the data for the three-qubit example are \{36, 13, 64, 71,
  14, 16, 7, 15, 60, 10, 84, 63, 64, 9, 55, 71, 8, 12, 10, 16, 16, 48, 67, 62,
  9, 64, 75, 63, 10, 74, 60, 73, 65, 14, 62, 66, 9, 57, 76, 53, 82, 78, 128,
  22, 61, 44, 25, 27, 56, 12, 52, 66, 14, 76, 56, 78, 45, 47, 22, 27, 66, 68,
  25, 102\}; and the data for the four-qubit example are \{5, 5, 11, 9, 5, 6,
  3, 5, 11, 4, 18, 18, 11, 3, 10, 24, 5, 4, 2, 5, 1, 5, 1, 2, 2, 1, 3, 1, 5, 6,
  3, 4, 16, 3, 24, 15, 3, 4, 4, 0, 16, 4, 27, 14, 13, 1, 17, 20, 9, 4, 16, 16,
  4, 2, 2, 4, 21, 3, 27, 23, 19, 4, 18, 35, 2, 3, 1, 6, 4, 0, 4, 5, 4, 6, 7, 4,
  3, 3, 1, 6, 0, 4, 4, 4, 0, 15, 14, 5, 2, 5, 19, 17, 3, 8, 14, 12, 2, 2, 2, 2,
  2, 16, 13, 13, 3, 26, 26, 21, 7, 27, 13, 13, 2, 1, 2, 4, 2, 8, 17, 21, 2, 23,
  21, 13, 4, 24, 21, 18, 7, 5, 17, 19, 4, 1, 9, 1, 10, 3, 25, 17, 19, 4, 15,
  18, 1, 6, 1, 1, 3, 15, 15, 17, 3, 20, 26, 21, 1, 15, 21, 17, 14, 5, 41, 16,
  10, 14, 40, 23, 35, 22, 71, 20, 14, 17, 22, 3, 13, 0, 16, 17, 1, 9, 12, 23,
  16, 17, 20, 1, 14, 27, 2, 17, 12, 3, 17, 20, 4, 6, 4, 3, 17, 4, 19, 13, 17,
  3, 14, 21, 5, 5, 4, 6, 3, 13, 11, 18, 1, 17, 20, 19, 2, 20, 19, 26, 21, 4,
  20, 21, 3, 24, 22, 14, 18, 17, 22, 4, 15, 17, 6, 14, 18, 6, 11, 21, 4, 24,
  18, 28, 18, 16, 0, 21, 30, 28, 11, 67\}.

\bibitem[Len \em{et~al.}(2018)Len, Dai, Eng\-lert, and Krivitsky]{Len2018}
Len, Y.L.; Dai, J.; Eng\-lert, B.G.; Krivitsky, L.A.
\newblock {Unambiguous path discrimination in a two-path interferometer}.
\newblock {\em Phys. Rev. A} {\bf 2018}, {\em 98},~022110.
\newblock
  doi:{\changeurlcolor{black}\href{https://doi.org/10.1103/PhysRevA.98.022110}{\detokenize{10.1103/PhysRevA.98.022110}}}.

\bibitem[Shang \em{et~al.}(2013)Shang, Ng, Sehrawat, Li, and
  Eng\-lert]{Shang2013}
Shang, J.; Ng, H.K.; Sehrawat, A.; Li, X.; Eng\-lert, B.G.
\newblock {Optimal error regions for quantum state estimation}.
\newblock {\em {New J. Phys.}} {\bf 2013}, {\em 15},~123026.
\newblock
  doi:{\changeurlcolor{black}\href{https://doi.org/10.1088/1367-2630/15/12/123026}{\detokenize{10.1088/1367-2630/15/12/123026}}}.

\bibitem[Nott \em{et~al.}(2021)Nott, Seah, Al-Labadi, Evans, Ng, and
  Eng\-lert]{Nott2021}
Nott, D.J.; Seah, M.; Al-Labadi, L.; Evans, M.; Ng, H.K.; Eng\-lert, B.G.
\newblock {Using prior expansions for prior-data conflict checking}.
\newblock {\em Bayesian Analysis} {\bf 2021}, {\em 16},~203--231.
\newblock
  doi:{\changeurlcolor{black}\href{https://doi.org/10.1214/20-BA1204}{\detokenize{10.1214/20-BA1204}}}.

\bibitem[Bera \em{et~al.}(2023)Bera, Bae, Hiesmayr, and
  Chru\'{s}ci\'{n}ski]{Hiesmayr+3:23}
Bera, A.; Bae, J.; Hiesmayr, B.C.; Chru\'{s}ci\'{n}ski, D.
\newblock On the structure of mirrored operators obtained from optimal
  entanglement witnesses.
\newblock {\em Sci. Rep.} {\bf 2023}, {\em 13},~2045--2322.
\newblock
  doi:{\changeurlcolor{black}\href{https://doi.org/10.1038/s41598-023-37771-0}{\detokenize{10.1038/s41598-023-37771-0}}}.

\bibitem[Liu and Chen(1998)]{Liu1998}
Liu, J.S.; Chen, R.
\newblock {Sequential Monte Carlo methods for dynamic systems}.
\newblock {\em J. Am. Stat. Assoc.} {\bf 1998}, {\em 93},~1032--1044.

\bibitem[Roberts and Rosenthal(2009)]{Roberts+1:09}
Roberts, G.O.; Rosenthal, J.S.
\newblock Examples of Adaptive MCMC.
\newblock {\em Journal of Computational and Graphical Statistics} {\bf 2009},
  {\em 18},~349--367.
\newblock
  doi:{\changeurlcolor{black}\href{https://doi.org/10.1198/jcgs.2009.06134}{\detokenize{10.1198/jcgs.2009.06134}}}.

\bibitem[Haario \em{et~al.}(2001)Haario, Saksman, and Tamminen]{Haario+2:01}
Haario, H.; Saksman, E.; Tamminen, J.
\newblock {An adaptive Metropolis algorithm}.
\newblock {\em Bernoulli} {\bf 2001}, {\em 7},~223 -- 242.

\bibitem[Drovandi and Pettitt(2011)]{MCstepNew}
Drovandi, C.C.; Pettitt, A.N.
\newblock {Estimation of parameters for macroparasite population evolution
  using approximate Bayesian computation}.
\newblock {\em Biometrics} {\bf 2011}, {\em 67},~225--233.

\bibitem[Roberts and Rosenthal(2001)]{MCar}
Roberts, G.O.; Rosenthal, J.S.
\newblock {Optimal scaling for various Metropolis--Hastings algorithms}.
\newblock {\em Stat. Sci.} {\bf 2001}, {\em 16},~351--367.

\end{thebibliography}
\end{document}